\documentstyle[psfig,epsf,oldgerm,graphics,psfrag]{mn}
\newcommand{\umin}{u_{\rm min}}
\newcommand{\te}{t_{\rm E}}

\newcommand{\dol}{D_{\rm OL}}
\newcommand{\dls}{D_{\rm LS}}
\newcommand{\dos}{D_{\rm OS}}
\newcommand{\re}{R_{\rm E}}
\newcommand{\rs}{R_{\rm s}}
\newcommand{\cs}{$\chi^{2}\:$} 

\newcommand{\Am}{$A_{\rm max}\:$}
\newcommand{\tf}{$t_{\rm FWHM}$}

\newcommand{\msol}{$M_{\odot}\:$}
\newcommand{\magmax}{m_{\rm max}}
\newcommand{\rspot}{R_{\rm spot}}

\begin{document}

\title{Planetary Microlensing at High Magnification}

\author[N.J. Rattenbury et al.]{N.J.~Rattenbury,$^1$ I.A.~Bond,$^{1,2}$ J.~Skuljan$^2$ and P.C.M.~Yock$^{1}$ \\
$^1$Department of Physics, University of Auckland, Auckland, New Zealand\\ 
$^2$Department of Physics and Astronomy, University of Canterbury, Christchurch, New Zealand
}
\date{Accepted 0000 January 00.
      Received 0000 December 00}

\pagerange{\pageref{firstpage}--\pageref{lastpage}}
\pubyear{2001}

\maketitle
\label{firstpage}
\begin{abstract}
Simulations of planetary microlensing at high magnification that were carried out on a cluster computer are presented. It was found that the perturbations due to two-thirds of all planets occur in the time interval $[-0.5t_{\rm FWHM},0.5t_{\rm FWHM}]$ with respect to the peak of the microlensing light curve, where $t_{\rm FWHM}$ is typically $\sim 14$ hours. This implies that only this restricted portion of the light curve need be intensively monitored for planets, a very significant practical advantage. Nearly all planetary detections in high magnification events will not involve caustic crossings. We discuss the issues involved in determining the planetary parameters in high magnification events. Earth mass planets may be detected with 1-m class telescopes if their projected orbital radii lie within about $1.5 - 2.5$ AU. Giant planets are detectable over a much larger region. For multi-planet systems the perturbations due to individual planets can be separated under certain conditions. The size of the source star needs to be determined independently, but the presence of spots on the source star is likely to be negligible, as is the effect of planetary motion during an event.
\end{abstract}

\begin{keywords}
Gravitational lensing: microlensing -- Methods: numerical -- Stars: planetary systems
\end{keywords}

\section{Introduction}

New results on extra-solar planets are now being obtained by various means, most notably the radial velocity, transit and gravitational microlensing techniques (Perryman, 2000). Gravitational microlensing events of high magnification have been shown to be promising in this regard (Griest \& Safizadeh, 1998; Rhie \emph{et al.}, 2000; Gaudi \emph{et al.}, 2002; Bond \emph{et al.}, 2002a). Using new observational and analytical techniques, it has recently been shown that high magnification events may be readily detected, including events with very high magnifications (Bond \emph{et al.}, 2002b). Here we discuss the detectability of planets in this new class of events using 1-m class telescopes, including:

\begin{itemize}

\item Critical observational period for planet detection
\item Zones of detectability of terrestrial and giant planets
\item Detectability of habitable planets and solar-system analogues.
\item Precision of planetary system characterisation
\item Multi-planet systems
\item Effect of the source star radius and stellar spots
\item Effect of orbital motion during a microlensing event
\item Comparison with low magnification events
\item Absolute characterisation of planetary parameters 

\end{itemize}

\section{Definitions used in this Study}
\label{sec:notation}
We consider the lensing of a source star by a lens star where the lens star is assumed to have a planet or planets orbiting it not dissimilar to those in our solar system. To first order, the lensing by such a system can be approximated by the lens star alone. In this case the lensing can be parametrised in terms of the Einstein radius $\re$ and crossing time $\te$ where

\begin{equation}
\re = 4.42 \sqrt{\frac{M_{\rm L}}{0.3M_{\sun}}} \sqrt{\frac{\dos}{8 \rm kpc}} \sqrt{\left(1 - d\right)d} \:\:\rm AU
\label{eq:bigre}
\end{equation}
and
\begin{equation}
\te = 38.25\sqrt{\frac{M_{\rm L}}{0.3M_{\sun}}} \sqrt{\frac{\dos}{8 \rm kpc}} \sqrt{\left(1 - d\right)d}\:\: \rm days.
\label{eq:bigte}
\end{equation}

$\dol$ and $\dls$ are the observer-lens and lens-source distances respectively and $d = \frac{\dol}{\dos}$. We see that $\re \simeq 1.9$ AU and $\te \simeq 16.6$ days for bulge events when the lens mass $M_{\rm L} \simeq 0.3M_{\odot}$, $\dol \simeq 6$kpc, $\dos \simeq 8$ kpc. The transverse velocity, $v_{\rm T}$, of the lens with respect to the observer-source line of sight is  taken as $200$ $\rm{kms^{-1}}$.

High magnification occurs when the lens and source stars are well aligned, i.e. when the impact parameter $\umin \ll \re$. In this case
\begin{equation}
A_{\rm max} \simeq \frac{\re}{\umin} \gg 1
\end{equation}
A useful quantity in high magnification events is the time of full-width-half-maximum with respect to the peak given by:
\begin{equation}
t_{\rm FWHM} = \frac{3.5\te}{A_{\rm max}} .
\label{eq:tfwhm}
\end{equation}
\section{General Structure and Critical Time of Planetary Deviations in High Magnification Events}
\label{sec:deviations}
Liebes (1964) discussed several features of gravitational microlensing events with magnifications up to $\sim 1000$, and remarked that planets orbiting the lens stars in these events would perturb the light curves. Griest \& Safizadeh (1998) subsequently computed the perturbations for a variety of high magnification configurations, and showed that the probability for detecting planets in these events is high. In particular, they showed that planets with masses as low as $10M_{\oplus}$ could be detected with significant probability in events with magnifications $\sim 50$ by monitoring the peaks of the events with a photometric precision $\sim 1\%$.

It has subsequently been shown that events with magnification up to $\sim 200$ are quite readily detectable, and also that intensive photometric monitoring of the peaks of these events with a precision significantly better than $1\%$ is feasible (Bond \emph{et al.}, 2002a; Bond \emph{et al.}, 2002b). Consequently the prospects for planet detection have improved. 

In this section we examine typical light curves for events of higher magnification, to determine the basic requirements for planet detection. The light curves presented here, and indeed throughout the paper, were generated using a rapid version of the inverse ray shooting method (Schneider \& Wei\ss, 1986; Kayser, Refsdal \& Stabell, 1986; Wambsganss, 1997). The technique was implemented on a $\sim 200$ node cluster computer at the University of Auckland (Dobcs\'anyi, 1999). A further improvement in the code runtime is described in Appendix A.

\begin{figure}
\psfrag{Planet at (xp,yp)}[][]{Planet at $(x_{\rm p},y_{\rm p})$}
\psfrag{Star at (0,umin)}{Star at $(0,u_{\rm min})$}
\psfrag{Source star track}{Source star track}
\psfrag{y}{y}
\psfrag{x}{x}
\epsfxsize=\hsize\epsfbox{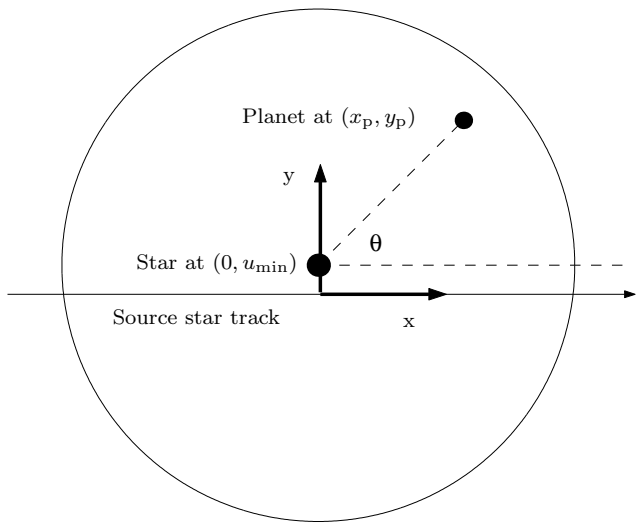}
\caption{\label{fig:coords}Co-ordinate system used by the inverse ray shooting code. The scales are in units of the Einstein ring radius, $\re$. In what follows the projected orbit radius at the peak of the microlensing, $\sqrt{x_{\rm p}^{2} + (y_{\rm p} - u_{\rm min})^{2}}$, is denoted by $a$.}
\end{figure}

In Figure \ref{fig:library} we show a selection of light curves for a typical lens system consisting of a star with a single planet in a typical event of high magnification. The fractional difference light curves are plotted: $\delta = \frac{\left(A_{\rm p} - A_{\rm s}\right)}{A_{\rm s}}$ where $A_{\rm p}$ is the light curve of the lens star including its planet and $A_{\rm s}$ is the light curve of the lens star alone. The planet-star mass ratio of the lens is $\epsilon = 4\times 10^{-5}$. This corresponds to a planet of four Earth masses orbiting a $0.3M_{\odot}$ mass star, the most probable lens mass. The maximum magnification is $A_{\rm max} = 80$. The radius of the source star, $R_{\rm s}$, has been set equal to $R_{\odot}$, as is done throughout this paper unless otherwise stated. The projected orbital radius of the planet at the time of the lensing event is $0.8\re$ or $\simeq 1.5$AU. The lens geometry is shown in Figure \ref{fig:coords}. 

It is clear from Figure \ref{fig:library} that the position angle of a planet at the time of lensing may be determined from the light curve. This follows because the peak deviation occurs at the point on the source star track when it is crossed by the planet-star axis. Thus the position angle $\theta$ is given by the equation
\begin{equation}
\tan \theta \simeq -\frac{\umin}{v_{\rm T} t_{\rm p}}
\label{eq:tp}
\end{equation}
where $t_{\rm p}$ denotes the time of maximum planetary deviation. Equation \ref{eq:tp} implies that, provided $\theta$ is not within $\pm 30^{\circ}$ of either $0^{\circ}$ or $180^{\circ}$, then $t_{\rm p}$  lies in the interval $[-0.5t_{\rm FWHM},0.5t_{\rm FWHM}]$. It follows that for two-thirds of all planetary positions, the perturbations they produce to the microlensing light curve occur in the very restricted time interval $[-0.5t_{\rm FWHM},0.5t_{\rm FWHM}]$. This is the critical time for planet searches.

\begin{figure}
\resizebox{0.5\textwidth}{0.5\textheight}{
\includegraphics{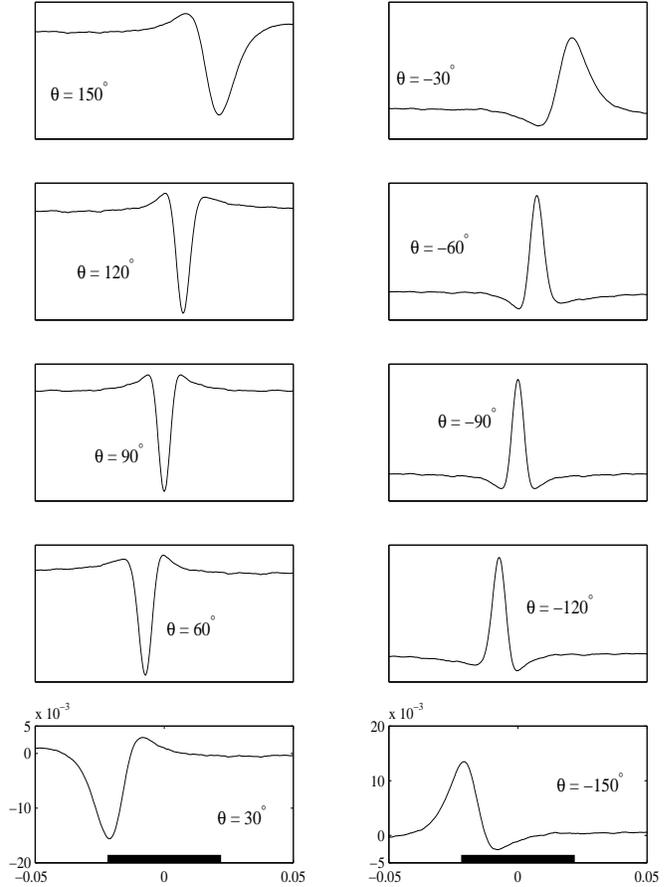}}

%\vspace{5.5cm}
\caption{\label{fig:library}Sample perturbed light curves for a single planet with mass ratio $\epsilon = 4\times 10^{-5}$. Each graph shows the deviation from the single lens light curve. The maximum light curve amplification is \Am = 80. The angle between the source star track and the planet-star axis is denoted by $\theta$ and defined in Figure \ref{fig:coords}. For negative values of $\theta$ the source star track crosses the planet-star axis between the planet and star. The projected planet orbit radius is $0.8\re$. For $\theta = 0^{\circ} \: \rm{and}\: 180^{\circ}$, there is a planetary caustic crossing, resulting in a large deviation from the single lens light curve. The time axis is in normalised time units, $t_{\rm N} = \frac{t}{t_{\rm E}}$. In these units $t_{\rm FWHM} = 0.044$ when $A_{\rm max} = 80$. The interval $[-0.5t_{\rm FWHM},0.5t_{\rm FWHM}]$ is indicated on the lower-most set of x-axes as a thick line.}
\end{figure}

\begin{figure}
\psfrag{xlabel}[][cB]{$t_{\rm N}$}
\epsfxsize=\hsize\epsfbox{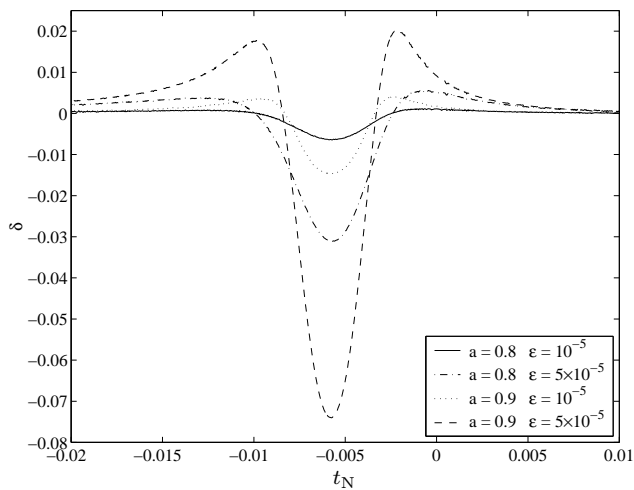}
%\vspace{5.5cm}
\caption{\label{fig:shapes}Light curve perturbations for various planet-star mass ratios and separations. The maximum magnification $A_{\rm max} = 100$, and $\theta = 60^{\circ}$.}
\end{figure}

Figure \ref{fig:shapes} shows typical examples of deviations produced by planets of different masses and at different projected radii. In general, heavier planets produce larger deviations, as do planets nearer the Einstein ring. The shapes of the deviations are also dependent on these properties. It is thus possible to determine both the mass and the projected radius of a planet from the deviation to the light curve. In Section \ref{sec:accuracy} we discuss the precision with which this may be achieved.

\section{Planet Detectability}
\label{sec:sensitivity}
Planetary zones of detectability in high magnification events were determined by comparing simulated perturbations to microlensing light curves produced by planets of various mass and at various projected radii with minimum detectable perturbations.

\subsection{Simulation details}
\label{sec:simdetails}
Each simulated planetary light curve was comprised of $301$ evenly spaced points in the critical time interval \tf. In order to simulate the observation process, noise was added to the simulated light curves. As a representative example, we assumed observations of 200s duration were made with a 1m telescope in a passband of 200nm. In this case, the number of detected photons per exposure is given by the expression (Eccles 1983):

\begin{equation}
N = \pi f\times 10^{(-0.4m +11.5)} \hspace{1cm} \rm{photons} 
\label{eq:flux}
\end{equation}
where $f$ denotes the telescope throughput. The number at any magnification, given a maximum magnification, \Am$\!$, is:
\begin{equation}
N(A) = N_{\rm {max}} \times \frac{A}{A_{\rm {max}}} \hspace{1cm} \rm{photons} .
\label{eq:noise1}
\end{equation}
If we assume Poisson statistics, the photon noise, $\sigma_{\rm N}$, is $[{N(A)}]^{-\frac{1}{2}}$. Using Equations \ref{eq:flux} and \ref{eq:noise1}, this becomes:
\begin{equation}
\sigma_{\rm N}(A) = \left[ \pi f \times 10^{(-0.4\magmax +11.5)} \frac{A}{A_{\rm{max}}} \right]^{-\frac{1}{2}}
\label{eq:noise2}
\end{equation}
where $\magmax$ is the magnitude at peak magnification. Equation \ref{eq:noise2} defines our noise function for any light curve with peak amplification \Am$\!$ and peak magnitude $m_{\rm max}$. The minimum error that the difference imaging technique can achieve is typically:
\begin{equation}
\sigma_{\rm DI}(A) = \sqrt{2}\sigma_{\rm N}(A)
\label{eq:DIlimit}
\end{equation}
because it involves the difference of two similar quantities. In what follows we assumed $f=0.3$ and Gaussian noise with $\sigma'_{\rm N} = 2\sigma_{\rm N}$. This implies $\sigma'_{\rm N} = 0.36\%$ at the peak of a light curve with $\magmax = 15$, consistent with the results reported by Bond (2002a).

\subsection{Results}
\begin{figure*}
\epsfxsize=\hsize\epsfbox{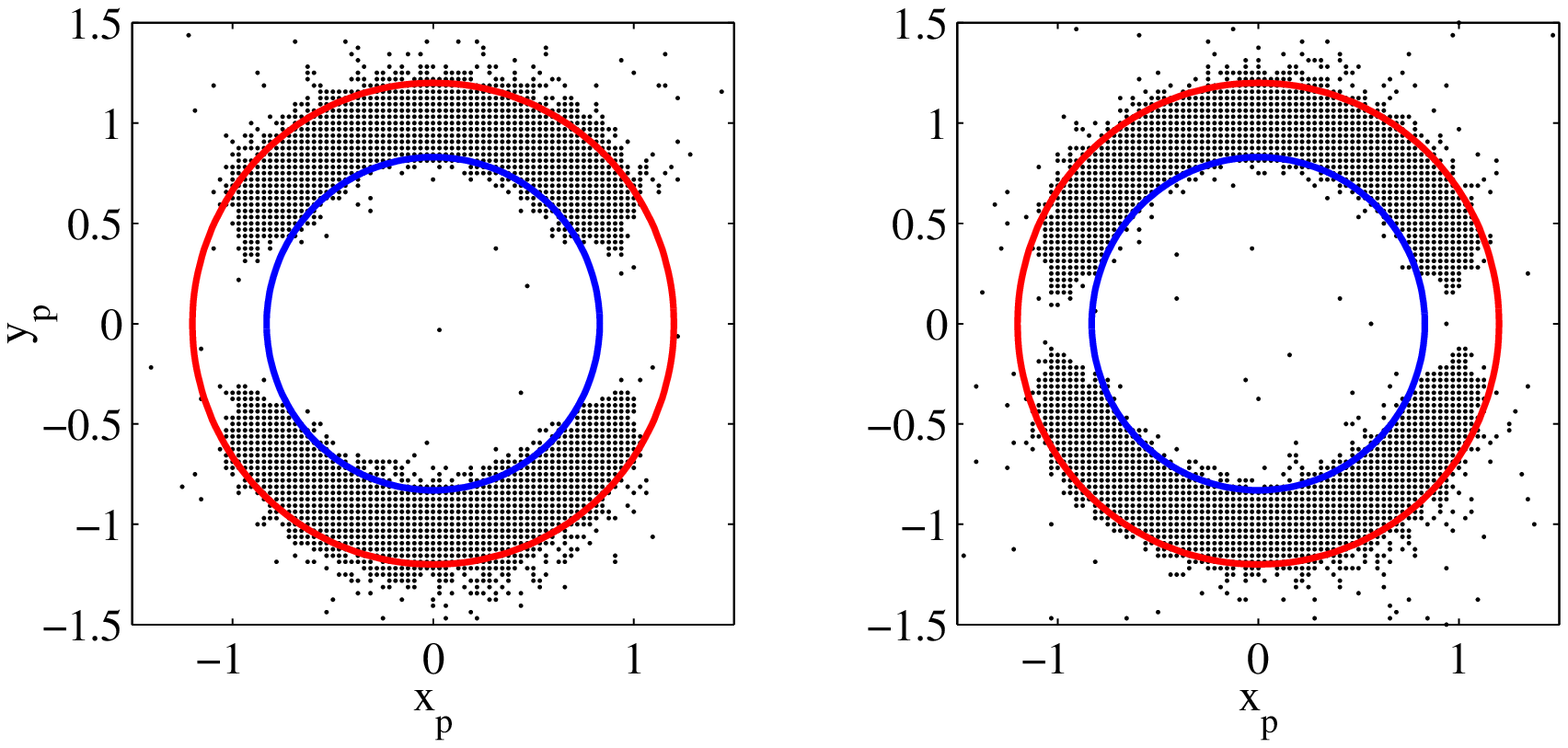}
%\vspace{6cm}
\caption{\label{fig:limits_ex}Detection limit maps for an Earth mass planet orbiting a $0.3M_{\odot}$ lens star. The axes are in units of $\re$. The I band magnitude at maximum is 15, and the maximum amplification is 100. The \cs map on the left was generated using observation light curves comprised of 301 points over the interval $[-\frac{1}{2}t_{\rm FWHM},\frac{1}{2}t_{\rm FWHM}]$. The right map was created using light curves comprised of 601 points over the interval $[-t_{\rm FWHM},t_{\rm FWHM}]$. The detection limits do not change greatly with increased observation interval, but the detection efficiency increases slightly.}
\end{figure*}

For each light curve \cs was computed using the single lens light curve. This was compared to $\chi^{2}_{0}$, the value obtained from data without any planetary signal. A deviation in the light curve was considered detectable if the difference between these two values, $\Delta \chi^{2} = \chi^{2} - \chi^{2}_{0}$, was greater than 60. Figure \ref{fig:limits_ex} shows the positions for an Earth-mass planet which give $\Delta \chi^{2} \geq 60$. The threshold value of 60 was chosen because it corresponds to $\sim 20$ consecutive deviations of 1-2$\sigma$ (Bond \emph{et al.}, 2002a). For events with $A_{\rm max} \sim 100$, the total number of measurements that can be made in the time interval $[-0.5t_{\rm FWHM}, 0.5t_{\rm FWHM}]$ is $\sim 300$ assuming parameters as above and a readout time between exposures of the CCD camera $\sim 100$ secs. If measurements are made over the interval $[-t_{\rm FWHM}, t_{\rm FWHM}]$, then $\sim 600$ measurements may be anticipated. The probability for a random fluctuation to yield $\Delta \chi^{2} \geq 60$ is $< 1\%$ for 300 measurements, and $\sim 4\%$ for 600 measurements. However, any random fluctuation will be easily identifiable in any densely sampled light curve, because it will not produce a coherent perturbation. The detection limits for the microlensing situations that were tested are shown in Table \ref{tab:detlims90_1}.

\begin{table*}
\caption{\label{tab:detlims90_1}Planet-star orbit detection limits. The planet mass ratios used were: $1\times 10^{-5},\; 1.7\times 10^{-4} \rm \:and\: 3.2\times 10^{-3}$. These values correspond to Earth, Neptune and Jupiter mass planets, given a 0.3\msol lens star. Each simulation light curve was comprised of 301 points over the interval: $[-\frac{1}{2}t_{\rm FWHM} , \frac{1}{2}t_{\rm FWHM}]$. Two values of maximum light curve magnitude are used, and three values of maximum light curve amplification. For a perturbation to be considered detectable, the difference in \cs between the perturbed light curve and a single lens light curve must exceed 60. The detection limits are in units of $\re$.}
\begin{tabular}{lcccccc}
\multicolumn{7}{c}{$M_{\rm L} = 0.3M_{\odot}$ $R_{\rm s} = R_{\odot}$} \\
\hline
\Am & \multicolumn{2}{c}{50} & \multicolumn{2}{c}{100} & \multicolumn{2}{c}{200} \\ \hline
$\magmax$ & 15 & 17 & 15 & 17 & 15 & 17 \\
Earth & 0.88 - 1.13 & 0.95 - 1.05 & 0.83 - 1.2 & 0.92 - 1.07 & 0.8 - 1.3 & 0.9 - 1.1 \\
Neptune & 0.5 - 2.0 & 0.7 - 1.6 & 0.4 -  & 0.6 - 1.9 & 0.1 - & 0.4 -  \\
Jupiter & 0.0 - & 0.1 - & 0.0 - & 0.0 - & 0.0 -  & 0.0 - \\ \hline
\multicolumn{7}{c}{}\\

\multicolumn{7}{c}{$M_{\rm L} = 0.3M_{\odot}$ $R_{\rm s} = 2R_{\odot}$} \\
\hline
\Am & \multicolumn{2}{c}{50} & \multicolumn{2}{c}{100} & \multicolumn{2}{c}{200} \\ \hline
$\magmax$ & 15 & 17 & 15 & 17 & 15 & 17 \\
Earth & 0.9 - 1.1 & 0.99 - 1.01 & 0.87 - 1.15 & 0.99 - 1.01 & 0.84 - 1.2 & 0.99 - 1.01 \\
Neptune & 0.5 - 2.0 & 0.6 - 1.5 & 0.4 -  & 0.5 - 2.0 & 0.1 - & 0.4 -  \\
Jupiter & 0.0 - & 0.1 - & 0.0 - & 0.0 - & 0.0 -  & 0.0 - \\ \hline
\multicolumn{7}{c}{}\\

\multicolumn{7}{c}{$M_{\rm L} = 0.3M_{\odot}$ $R_{\rm s} = 4R_{\odot}$} \\
\hline
\Am & \multicolumn{2}{c}{50} & \multicolumn{2}{c}{100} & \multicolumn{2}{c}{200} \\ \hline
$\magmax$ & 15 & 17 & 15 & 17 & 15 & 17 \\
Earth & 0.94 - 1.06 & -  & 0.94 - 1.06 &  - & 0.99 - 1.01 & -  \\
Neptune & 0.5 - & 0.65 - 1.54 & 0.3 -  & 0.5 - 2.0 & 0.2 - & 0.3 -  \\
Jupiter & 0.0 - & 0.1 - & 0.0 - & 0.0 - & 0.0 -  & 0.0 - \\ \hline
\multicolumn{7}{c}{}\\

\multicolumn{7}{c}{$M_{\rm L} = M_{\odot}$ $R_{\rm s} = R_{\odot}$} \\
\hline
\Am & \multicolumn{2}{c}{50} & \multicolumn{2}{c}{100} & \multicolumn{2}{c}{200} \\ \hline
$\magmax$ & 15 & 17 & 15 & 17 & 15 & 17 \\
Earth & 0.93 - 1.08 & 0.94 - 1.06 & 0.91 - 1.1 &  0.92 - 1.09 & 0.88 - 1.14 & 0.89 - 1.12 \\
Neptune & 0.6 - 1.6 & 0.75 - 1.3 & 0.5 - 2.0 & 0.5 - 2.0 & 0.4 - & 0.4 -  \\
Jupiter & 0.0 - & 0.0 - & 0.0 - & 0.0 - & 0.0 -  & 0.0 - \\ \hline
\end{tabular}
\end{table*}

The trends in the results are clear. The detection limits extend around the Einstein radius in the lens plane with heavier planets detectable further away from this region. 

\subsection{Habitable planets}

Numerical simulations by Wetherill (1996) predict that $5 - 15\%$ of all stars with mass $0.5 - 1.5M_{\odot}$ have a habitable planet. We carried out simulations to determine if this interesting prediction could be tested in microlensing events of high magnification. Following Kasting (1993) and Wetherill (1996), and allowing for internal heat sources such as long-lived radionuclides (Perryman, 2000), we assumed -- see Table \ref{tab:HZs} -- the following limits for the habitable zone around a star of approximately solar mass: $0.8 - 2$ AU. Following Deeg (2000), the mass range of habitable planets was taken to be $0.8 - 3M_{\oplus}$, although a broader range could have been assumed (Perryman, 2000).

\begin{table}
\caption{\label{tab:HZs}Habitable Zone boundaries during the first $10^{9}$ years (Wetherill, 1996; Kasting, 1993).}
\begin{tabular}{cc}
Stellar Mass $(M_{\odot})$ & HZ boundaries $(AU)$\\
0.5 & 0.20 - 0.40 \\
1.0 & 0.82 - 1.40 \\
1.5 & 1.70 - 2.80 \\
\end{tabular}
\end{table}

\begin{figure}
\epsfxsize=\hsize\epsfbox{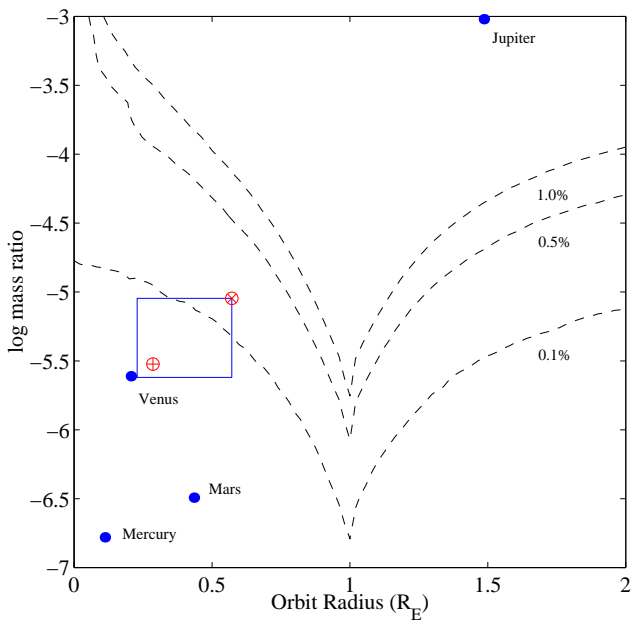}
%\vspace{5.5cm}
\caption{\label{fig:limitsHZ}Detection limit contours with the HZ (see text). The parameter positions used in the simulations of both an Earth and a heavy, cool Earth are denoted by the $\oplus$ and $\otimes$ symbols respectively  The parameter values of other solar system planets are indicated also. The mass ratio and Einstein ring radii are scaled for a solar mass lens star. The simulated light curves were computed over the time interval $[-2t_{\rm FWHM},2t_{\rm FWHM}]$, with $A_{\rm max} = 100$.}
\end{figure}

Simulated light curves for planets over a wide range of masses and orbital radii were generated. Each simulated light curve was compared with its associated singe lens light curve, and the maximum deviation, $\delta_{\rm max}$, was calculated. These maximum deviations were plotted on the mass-orbit plane, see Figure \ref{fig:limitsHZ}. Contours for $\delta_{\rm max} = 0.1\% , 0.5\%$ and $1.0\%$ are shown. In practice, the minimum detectable value for $\delta_{\rm max}$ for a 1-m class telescope will lie somewhere between the $0.1\%$ and $0.5\%$ contours, with the lower limit being more closely achieved by a space-borne telescope. We thus conclude that, whilst terrestrial planets are certainly detectable in microlensing events of high magnification, habitable planets are only marginally detectable, and Earth-Sun analogues are probably not detectable using this technique with 1m class telescopes.

\section{Precision of planetary system characterisation}
\label{sec:accuracy}

\begin{figure}
\epsfxsize=\hsize\epsfbox{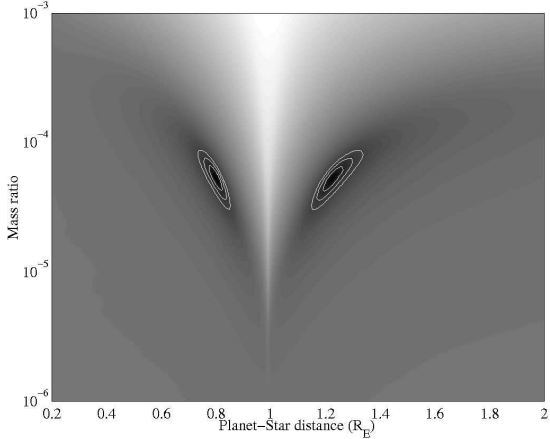}
%\vspace{6cm}
\caption{\label{fig:degen1}\cs computed on the planet distance-mass plane for a planet at $(a,\epsilon) = (0.8\re,5\times 10^{-5})$.  The angle that the planet-star axis makes with the source star track is 30 degrees and $A_{\rm max} = 100$. The contours indicated in the figure are the 1, 2 and 3$\sigma$ confidence intervals, corresponding to $\Delta \chi^{2} = $ 2.3, 6.17 and 11.8 respectively.}
\end{figure}

To compute the precision at which planetary masses and orbital radii may be measured, a single light curve with a planetary deviation was compared to a set of model light curves where the planet position and mass were varied over suitable ranges for a typical high magnification event with $A_{\rm max} = 100$.  As in previous analyses, a \cs degree of merit was computed for each comparison curve, assuming measurements were made at 5 minute intervals, as above. A typical plot of the \cs values over the mass-position plane for a planet at $(a,\epsilon) = (0.8\re,5\times 10^{-5})$ is shown in Figure \ref{fig:degen1}. Two \cs minima are visible, one at $(a,\epsilon) = (0.8\re,5\times 10^{-5})$, which corresponds to the light curve under analysis itself, and one at the same mass value, but with $a = 1.25\re$. This is the well known degeneracy in planet-star distance, where a planet at orbital distance $a$ produces a light curve perturbation which is indistinguishable from that with a planet at distance $1/a$ (Griest \& Safizadeh, 1998). There are no other \cs minima seen in the mass-position plane, indicating that a low mass planet near the $\re$ cannot be modelled by a higher mass planet further away. Further simulations using other planet mass-position values show that the confidence intervals for planet mass and position increase for a planet closer to the Einstein ring. Typically, the precision of mass measurements for terrestrial planets will be $\pm 50\% $, and their instantaneous projected orbital radii will be determined to an accuracy of typically $\pm 5\%$. The latter measurement will be subject to further uncertainty due to the degeneracy mentioned above.

\section{Multi-planetary Systems}
\label{sec:multiplanet}

Gaudi \emph{et al.} (1998) commented on the need to consider the effects of multi-planet lens systems in high magnification events. However, even an initial search for a single planet signal in a light curve involves a large amount of computation. An initial search for a two planet system which varies the minimum number of parameters would require $\mathcal{O}$$(10^{11})$ simulations. Clearly, searching the parameter spaces for a multi-planet system is impossible without \emph{a priori} information. 

Rattenbury (2001) and Bond \emph{et al.} (2002a) suggested that if an observed light curve shows evidence of perturbations due to a multi-planet lens system, it may not be necessary to search the planet position and mass parameter spaces completely in order to identify the lens system components. Rather, the more complicated lens system can be approximated as a combination of single planet systems. A similar result was subsequently reported by Han \emph{et al.} (2001) for low magnification events. Here we investigate under what circumstances the result is valid in high magnification events.

\begin{figure}
\epsfxsize=\hsize\epsfbox{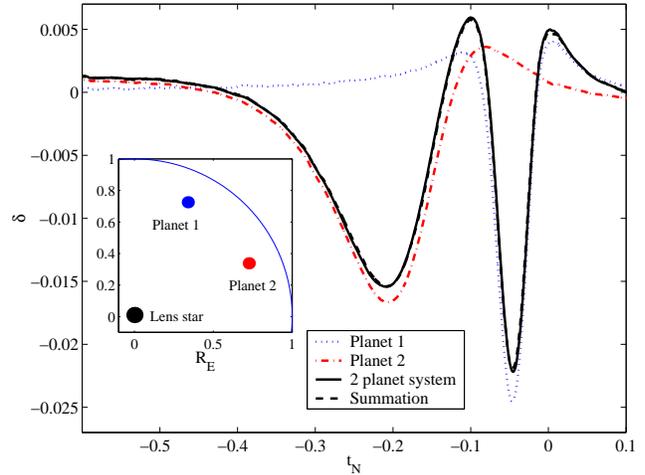}
\caption{\label{fig:2planet}$\delta$ light curves showing that the synthesis of two single planet light curves is an adequate first approximation to a two planet system. On this scale, the ``two-planet'' and ``summation'' graphs are indistinguishable. In this example, the planets are of equal mass $(\epsilon_{1,2} = 4\times 10^{-5})$, each at $0.8 \re$ from the lens star with a separation of $40^{\circ}$. It is unlikely that two planets will have equal, or similar orbit radii, however they can have equal projected radii.}
\end{figure}

A series of two planet system light curves was generated, varying the angular separation between the two planets. The two planet masses were equal. We then combined the two light curves obtained using the corresponding single planet systems. The two planet system light curves and the light curves from synthesising single planet light curves were then compared. It was found that there was little difference between these light curves. Figure \ref{fig:2planet} shows two single planet curves, $S_{\rm 1}$, $S_{\rm 2}$; the actual two-planet light curve and the synthetic light curve $\hat{S} = S_{\rm 1} + S_{\rm 2}$.

\begin{figure}
\epsfxsize=\hsize\epsfbox{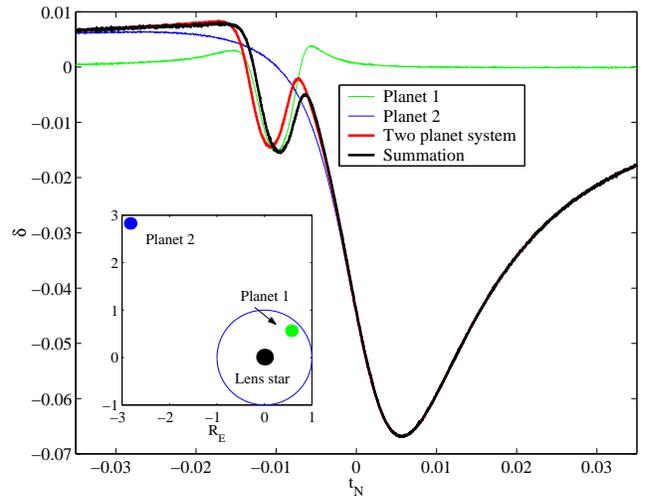}
\caption{\label{fig:jovtel}Synthetic (summation of single planet light curves) and multi-planet light curves for a two planet system. The planet parameters used are $\epsilon_{1} = 1\times 10^{-5}$, $\epsilon_{2} = 3\times 10^{-3}$; $a_{1} = 0.9\re$ and $a_{2} = 4\re$. These parameters correspond to Earth and Jupiter mass planets at projected radii 1.7 and 7.6 AU respectively, assuming a $0.3M_{\odot}$ mass lens star. $A_{\rm max} = 100$. The planet positions are indicated in the figure insert; $\theta_{1} = 45^{\circ}$ and $\theta_{2} = 135^{\circ}$.}
\end{figure}

A second two planet system was then modelled by a summation of the individual single planet light curves, see Figure \ref{fig:jovtel}. In this case the two planet masses were unequal, and a similar result was obtained. A three planet model was also tested in the same manner, with similar results. The above investigations confirm that multi-planet systems may indeed be approximated by a combination of single planet systems as a first approximation for more accurate modelling.

A further consideration in multi-planet systems is the limits that can be set on the resolution of separate planetary signals. By taking a series of two planet light curves with varying angular separation between the planets, we found that the separate perturbations can typically be resolved when the planets are separated by $\geq 20$ degrees. 

\section{Source Star Effects}
\subsection{Source size effects}
\label{subsec:source_effects}
\begin{figure}
\epsfxsize=\hsize\epsfbox{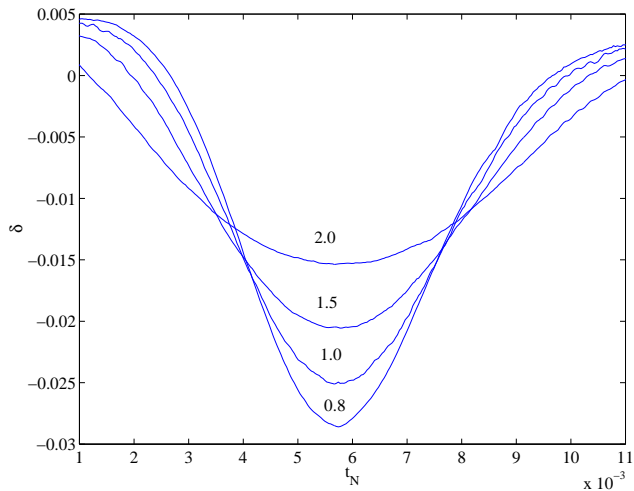}
\caption{\label{fig:sources}Effect of source star radius on a planetary perturbation. The units given are in solar radii, assuming a $0.3M_{\odot}$ lens star. The planet parameters are $(a,\epsilon) = (0.8\re,4\times 10^{-5})$ with $A_{\rm max} = 100$ and $\theta = 120^{\circ}$. }
\end{figure}

The effect of the size of the source star in planetary microlensing was first investigated by Bennett \& Rhie (1996). These authors considered events of low magnification, and found a significant effect. Subsequently Griest \& Safizadeh (1998), Rhie \emph{et al.} (2000) and Gaudi \emph{et al.} (2002) noted that a similar effect is present in events of high magnification. The effect is particularly amenable to investigation using the inverse ray shooting technique. Figure \ref{fig:sources} shows that as the source star size increases, the perturbation becomes smaller and broader. Comparison with Figure \ref{fig:shapes} shows that that variations of a planet's mass and projected orbital radius can produce qualitatively similar results. This is possibly the major shortcoming of the microlensing technique. Unless independent information is available to pin down the radius of a source star to $\sim 20\%$ accuracy, then the precision with which the mass and the projected orbital radius of a planet can be determined, as given in Section \ref{sec:accuracy}, will be significantly compromised. The achromaticity of gravitational lensing may be used to advantage in this regard, to obtain high quality photometry in several passbands of a source star when it is highly magnified. These measurements, combined with detailed knowledge of reddening along the line of sight, should suffice to determine the radius of the source star with adequate precision. 

\subsection{Source star spots}
The use of microlensing for the detection of small scale structure on the source star was recently suggested by Heyrovsk\'{y} \& Sasselov (2000). A non-uniform surface brightness distribution on the source star will produce amplification deviations from a source star with a uniform brightness distribution. These deviations can reach a significant level $(>2\%)$ for source-transit events (Heyrovsk\'{y} \& Sasselov, 2000). It was noted that light curve deviations due to source star spots may be mistaken for the perturbations due to low mass planetary companions of lens star. However Han \emph{et al.} (2000) notes that even if the source star transits the lens, the source star spot must pass very close to the lens before a detectable deviation will occur. Source star spots will therefore affect only a very small number of microlensing events.

In order to confirm that the perturbations due to source star spots are indeed negligible compared to those due to lens system planets, we performed a number of numerical simulations for typical microlensing geometries. Most microlensing events are not source transit events: the minimum impact parameter is many times the source star radius. The simulations were performed using a range of $\umin$, $\rs$ and $\rspot$ values, where $\rspot$ is the radius of a source star spot. The spot was moved to several locations on the source star and the brightness contrast was set at $+10\%$ with respect to the rest of the photosphere. Each light curve using a spotted source star was compared to a spotless source star. The deviation light curve,  $\delta_{\rm sp} = \frac{(A_{\rm sp} - A_{\rm s})}{A_{\rm s}}$, was computed, where $A_{\rm sp}$ and $A_{\rm s}$ are the amplification light curves of a spotted and spotless source star respectively. For each simulation, the maximum value of $|\delta|$ was recorded. The results are shown in Table \ref{tab:spots}. As we can see from the table, only in the most extreme case might the perturbation due to a source star spot approach the magnitude due to a low mass planet.

\begin{table}
\caption{\label{tab:spots}Table of source star spot parameters and maximum light curve deviation, $\max(|\delta_{\rm sp}|)$. These results are for a spot located at the centre of the source star disk.}
\begin{tabular}{cccc}
$\rs$ & \Am & $\rspot$ & $\max(|\delta_{\rm sp}|)$ \\
$R_{\odot}$ & 100 & $0.1\rs$ & $6\times 10^{-6}$ \\
$R_{\odot}$ & 100 & $0.5\rs$ & $1.3\times 10^{-4}$ \\
$2R_{\odot}$ & 200 & $0.1\rs$ & $9.1\times 10^{-5}$ \\
$2R_{\odot}$ & 200 & $0.5\rs$ & $2\times 10^{-3}$ \\
\end{tabular}
\end{table}

\subsection{Distance and other effects}
In the simulations presented here the parameters $\dos$, $\dol$ and $M_{\rm L}$ have been set at default values of 8 kpc, 6 kpc and $0.3 M_{\odot}$ respectively unless otherwise stated. These are typical values for real events, but they will be subject to large event-to-event variations in real cases. A complete analysis of the data for actual events will require direct knowledge of these parameters. Also, observations of an event during the critical time $t_{\rm FWHM}$ only, although sensitive to the presence of planets, will not suffice to determine accurately the value of \Am for an event. Knowledge of this quantity is required to quantify planetary parameters.

Additional measurements will be needed in actual events to determine the values of the above parameters. The distance to the source star $\dos$ may be determined approximately from the measurements described above for determining its radius. The remaining quantities may be evaluated by directly observing the lens and source stars a few years after a microlensing event, when they have diverged sufficiently to be resolved (Alcock \emph{et al}, 2001). Multi-passband photometry of the lens star would enable its type and mass $M_{\rm L}$ to be determined, and also its distance $\dol$ approximately. At the same time, the baseline intensity of the source star could be accurately measured, and \Am evaluated accurately. These measurements would provide absolute values of all microlensing parameters of the event. However, the value of the quantity $(1-d)$ that appears in Equation \ref{eq:bigre} would be quite uncertain, as it involves the difference of two similar but only approximately determined quantities, $\dos$ and $\dol$. This quantity could be better constrained through direct measurement of $v_{\rm T}$ and use of the equation $v_{\rm T} = \re/\te$. The net result of these measurements would be the absolute determination of the mass and projected orbital radius of any detected planet, and also the identification of its host star type. We note, however, that a telescope with capabilities approaching that of the VLT interferometer may be required.

\begin{figure}
\psfrag{u}{\Large u}
\psfrag{b}{\Large b}
\psfrag{lens}{\Large Lens}
\psfrag{source}{\Large Source}
\psfrag{earth}{\Large Earth}
\psfrag{satellite}{\Large Satellite}
\centering
\vspace{0.5cm}

\scalebox{0.7}{
\epsfxsize=\hsize\epsfbox{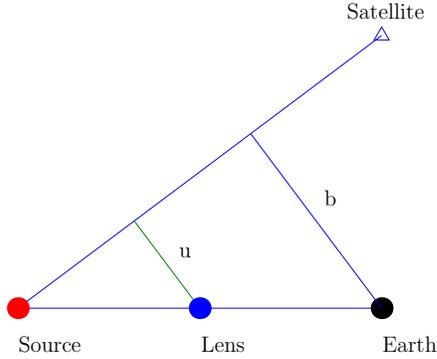}}
\caption{\label{fig:parallax}Parallax measurement for a high magnification event using a satellite in solar orbit. At the time of peak magnification, the source, lens and Earth may be treated as collinear.}
\end{figure}

Additional information on the parameters $\dos$, $\dol$, and $M_{\rm L}$ may be obtained through parallax measurements (Refsdal, 1966; Gould, 1995; Holz \& Wald, 1996). These would be most useful if carried out from a satellite in a distant (solar) orbit. As depicted in Figure \ref{fig:parallax}, the source, lens and Earth may then be treated as collinear at the time of peak magnification for a high magnification event. The amplification as detected by the satellite at this time could be determined from the light curve recorded by the satellite. This would yield the impact parameter $u$ of the lens at this time as seen from the satellite relative to the Einstein ring radius of the lens $\re$. Also, the baseline $b$ of the satellite could be determined from the satellite's orbit. These quantities would then satisfy the parallax equation:

\begin{equation}
\label{eq:parallax}
\frac{u}{b} = \frac{\dls}{\dos}.
\end{equation}

Equation \ref{eq:parallax} could be used to constrain the parameters $\dos$, $\dol$ and $M_{\rm L}$.

\section{Orbital motion during an event}
\begin{figure}
\psfrag{Deltaphi}{$\Delta\theta$}
\psfrag{a}{$a$}
\psfrag{RE}{$R_{\rm E}$}

\epsfxsize=\hsize\epsfbox{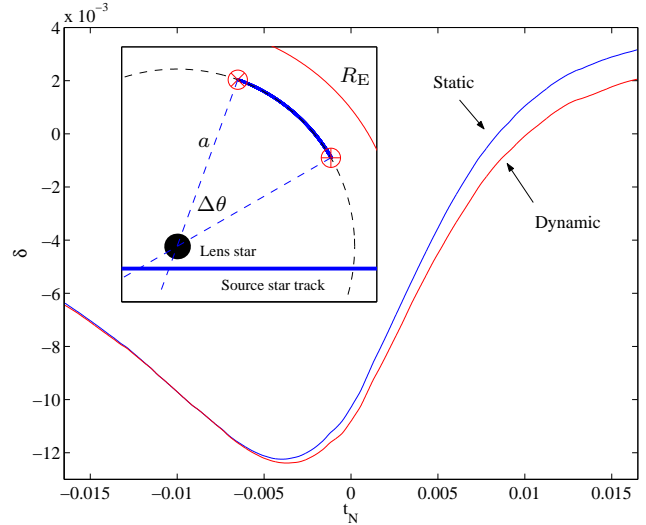}
\caption{\label{fig:dynshow}Planetary perturbation broadening due to a ``hot Jupiter'' moving in its orbit during an event. The intersection of the planet-star axis with the source star track is moving in the same sense as the source star, from left to right. The inset shows the dynamic microlensing geometry (not to scale). Using the simulation parameters of the text, the angle through which the planet moves during $t_{\rm FWHM}$ is $\Delta\theta \simeq 10.7^{\circ}$. The initial value of $\theta$ was $45^{\circ}$.}
\end{figure}

Dominik (1998) first studied the effect of orbital motion of a binary lens during a microlensing event. It is generally held that for most planetary microlensing events, the effect of the orbital motion of the planet around the lens star will be negligible. This assumption appears particularly safe for high magnification events, where the critical observation time \tf\, is so short.  To confirm this we determined typical parameters for which planetary motion during the peak of an event would be detectable. A face-on planetary system was assumed with a $0.3M_{\odot}$ lens star and a Jupiter mass planet, $\epsilon = 3.2\times10^{-3}$ orbiting at $0.05\re$. These parameters correspond to a ``hot Jupiter'' orbiting a typical lens star. The light curve shows a clear deviation from a model with a static planet, see Figure \ref{fig:dynshow}. The light curve perturbation due to a moving planet is seen to be broader (narrower) than that for a static planet when the direction of the source star is in the same (opposite) as the moving intersection of the planet-star axis with the source star track. The simulation was repeated for a Jupiter mass planet at $0.8\re$, using $v_{\rm T} = 20 \rm {kms^{-1}}$ and $A_{\rm max} = 10$. Similar results were obtained, with the broadening effect apparent during the whole peak perturbation.

\section{Comparison of the High magnification and caustic crossing techniques}

\label{sec:caustics}

The first planned searches for planets using the microlensing technique were not based on the high magnification technique that has been described here (Albrow \emph{et al.}, 1996; Pratt \emph{et al.}, 1996). Rather, they were based on the well-known ``caustic crossing'' technique (Mao \& Paczy\'{n}ski, 1991; Gould \& Loeb, 1992; Bolatto \& Falco, 1994; Bennett \& Rhie, 1996; Wambsganss, 1997; Bozza, 1999). Caustic crossing events, in which the light curve undergoes large, rapid changes in amplification, can yield a great deal of information about the lens system and the source star. However, even if a planet is known to be in the lens system, the probability for a caustic crossing to occur is low. Moreover, the time of occurrence of the caustic crossing is unpredictable. For these reasons, the observational study of planets by the caustic crossing technique has proved to be difficult. As a practical technique, it is preferable to be able to concentrate telescope resources during a short well-defined time period, rather than over a longer, indeterminate period. This is especially true for ground-based telescopes that are being used for several purposes.

Ultimately, a wide-angle space-borne telescope dedicated to microlensing could achieve the greatest sensitivity, by continuously monitoring a very large number of main sequence stars. This would yield sensitivity to terrestrial planets over a significantly wider range of orbital radii than the high magnification technique can achieve. The Galactic Exoplanet Survey Telescope (GEST) has been proposed to exploit this capability (Bennett \& Rhie, 2002).

\begin{figure}
\epsfxsize=\hsize\epsfbox{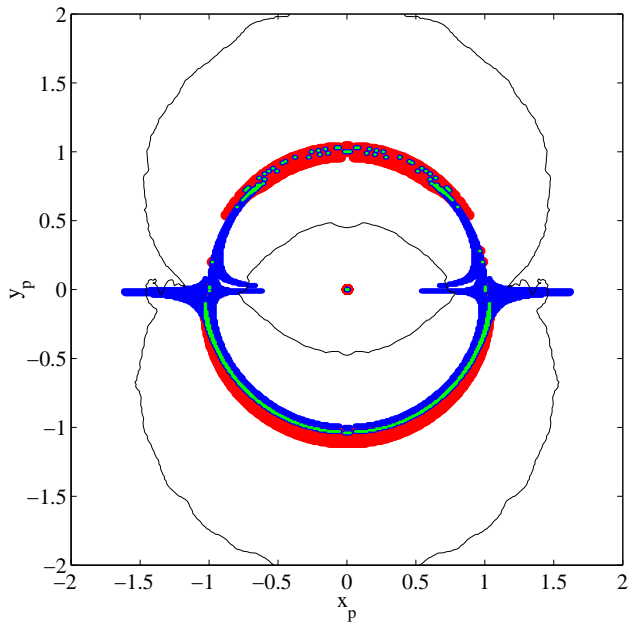}
\caption{\label{fig:caustic_map}Planet positions for which a caustic crossing occurs within $t \in [-\te,\te]$. The planet mass ratio for these planet positions is $\epsilon = 10^{-4}$ and the minimum impact parameter is $u_{\rm min} = 0.01$. The position axes are in units of $\re$. Planet positions where the source star track crosses the central caustic are plotted in grey; those where the track crosses a planetary caustic, with dark shading. Lightly shaded regions denote planet positions where both a planetary and central caustic crossing occurs. The solid line indicates the regions where a planet of the same mass can be detected via the high magnification technique, assuming a conservative detection accuracy of 1.0\%.}
\end{figure}

In order to further compare the caustic crossing and high magnification techniques, we estimated relative frequencies of planetary detections that may occur with and without caustic crossings in events of high magnification. From the perturbative analysis of Bozza (1999), the caustic curve maps for a range of single planet lens systems were generated, varying planet position, mass and the minimum impact parameter, $\umin$. For each set of planet parameters, it was determined if a caustic crossing occurred at any time within the interval $t \in [-\te,\te]$.  A solar radius source star was used. In order to preserve the perturbative nature of the analysis of Bozza, parameter sets where the first order perturbative term exceeded 0.05 were ignored. An example image of the results obtained in this analysis is shown in Figure \ref{fig:caustic_map}. The positions where the source star track crosses a planetary and/or the central caustic are shown. From this it is evident that nearly all planetary detections in high magnification events will not involve caustic crossings. 

\section{Conclusion}

The investigations presented here provide further support for the practicality of studying planetary systems in gravitational microlensing events of high magnification, including planets with masses as low as that of Earth. 

It has been shown that the perturbations produced by planets with position angles $30^{\circ} < \theta < 150^{\circ}$ or $210^{\circ} < \theta < 330^{\circ}$ with respect to the source star track occur in the restricted time interval $[-0.5t_{\rm FWHM},0.5t_{\rm FWHM}]$ about the peak of the light curve. Here, $t_{\rm FWHM} = \frac{3.5t_{\rm E}}{A_{\rm max}}$, and is typically $\sim 14$ hours when $A_{\rm max} \sim 100$. This implies that the planetary deviations of two thirds of all planets occur in this relatively restricted time interval, a significant advantage for carrying out practical observations. 

It has also been demonstrated that the position angle of a planet is simply related to the time and sign of the perturbation it produces on the light curve, and that its mass and projected orbit radius may be determined from the height and shape of the planetary deviation, provided the radius of the source star can be determined independently.     

It has been shown that Earth-mass planets are typically detectable with 1-m class telescopes provided they lie within about $25\%$ of the Einstein ring radius of the lens star, i.e. from about 1.5 - 2.5 AU. Neptune-mass are typically detectable provided they lie beyond about 0.8 AU, and Jupiter-mass planets are detectable almost anywhere. These zones of detectability, whilst large, only marginally include planets in the ``habitable zone''. It was also shown that nearly all planetary detections in high magnification events will not involve caustic crossings.

Multi-planet systems can be systematically characterised. To a good approximation, the joint perturbation is the sum of the individual perturbations of the component planets. This approximation may be used as a first approximation in planetary modelling.  

Source star effects have been investigated. The main effect is the size of the source star. This may be able to be determined by multi-passband photometry carried out during the course of a microlensing event. The effects of spots on the source star are likely to be negligible. The effect of orbital motion of a planet during a microlensing event has also been investigated. In most cases this is entirely negligible, although in the case of hot Jupiters it is likely to be significant.  

Follow-up observations carried out a few years after a microlensing event when the lens and source stars are resolvable should suffice to characterize the parameters of any detected planet, and its host star, absolutely.

Finally, we have found that the numerical inverse ray shooting technique for analysing microlensing events can be conveniently carried out on a cluster computer, and that complicated physical situations can thereby be rapidly simulated. 

\section{Acknowledgements}
The authors thank Peter Dobcs\'anyi for assistance, Bohdan Paczy\'{n}ski and Dave Bennett for comments, and the Marsden Fund of NZ for financial assistance.

\label{lastpage}
\appendix
\section{Inverse Ray Shooting Code Details}
\begin{figure}

\psfrag{B3}{$B_{m}$}
\psfrag{B2}{$B'_{m}$}
\psfrag{B1}{$\widehat{B}'_{m}$}
\psfrag{xlabel}{$x_{\rm p}$ (AU)}
\psfrag{ylabel}{$y_{\rm p}$ (AU)}

\epsfxsize=\hsize\epsfbox{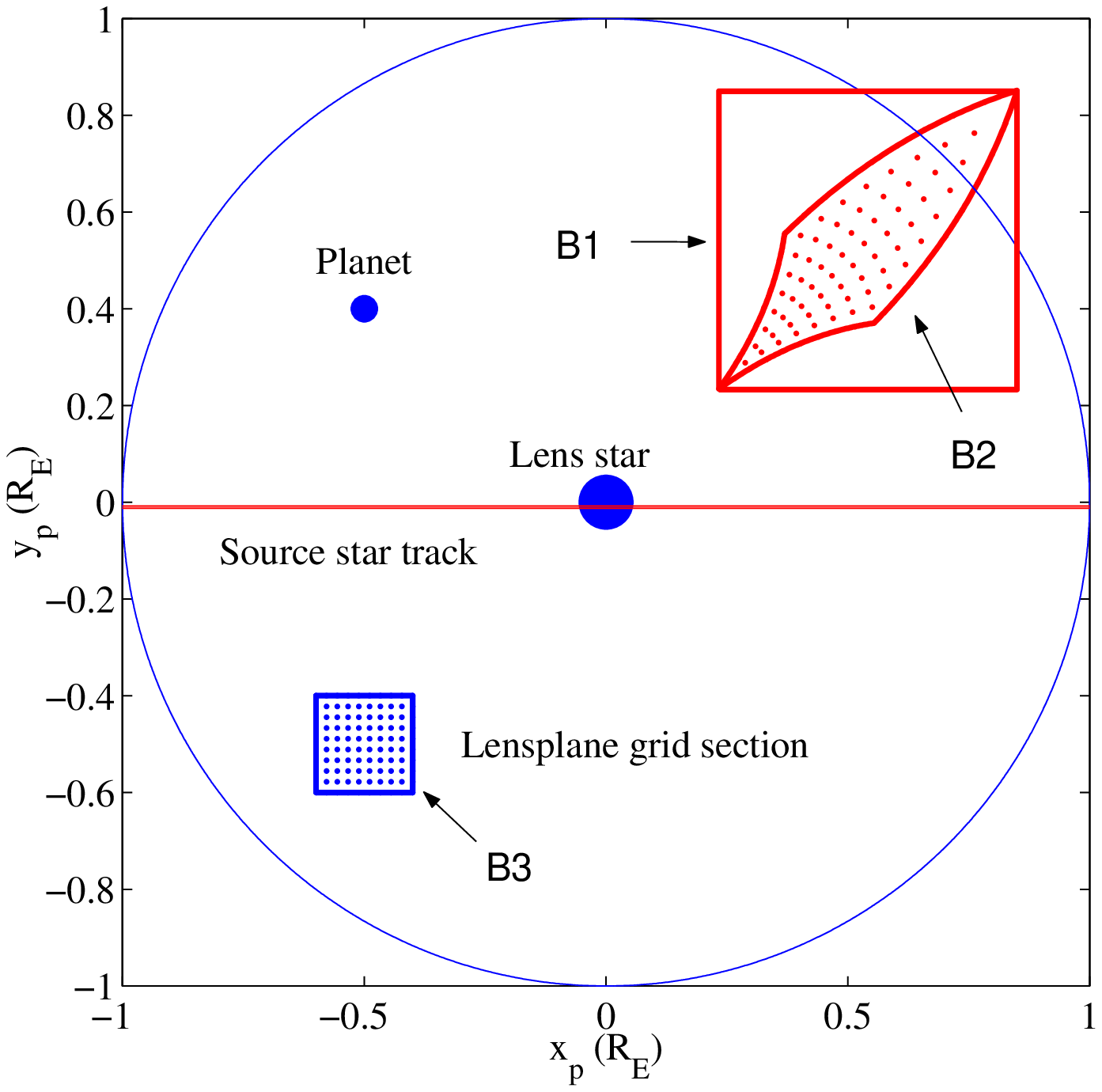}
\caption{\label{fig:map}Boundary box theorem: Any lens plane point $\bmath{L}$ contained within boundary box $\bmath{B_{m}}$, will be mapped to a point $\bmath{S}$ contained within box $\bmath{\widehat{B}_{m}'}$.}
\end{figure}

An improved version of the inverse ray shooting code has been developed which utilises a special property of the microlensing mapping functions. A rectangular region $\bmath{B_{m}}$ is mapped to a new region $\bmath{B'_{m}}$ in the source plane, see Figure \ref{fig:map}. It can be shown that for any lensplane point $\bmath{L} = (x,y)$ contained in a lensplane boundary box $\bmath{B_{m}}$, the corresponding source plane point $\bmath{S} = (\xi(x,y),\eta(x,y))$ will be contained within the boundary $\bmath{\widehat{B}'_{m}}$ which is the smallest rectangle in the source plane containing $\bmath{B'_{m}}$. The proof of this statement follows by demonstrating that, for regions $\bmath{B_{m}}$ not containing any lenses, the mapping functions $\xi$ and $\eta$ do not have any local extrema. If none of the elements of $\bmath{\widehat{B}'_{m}}$ are contained within the source star track region of interest, the area of the lensplane defined by $\bmath{B_{m}}$ can be discarded. The improved version of the code has a run rate almost one order of magnitude less than the original version. This is significant when modelling actual datasets.

\subsection*{Microlensing mapping functions}

Due to the gravitational microlensing, any given point $\bmath{L}(x,y)$
from the lens plane is mapped into a single point
$\bmath{S}(\xi,\eta)$ in the source plane:
 \begin{equation}
    \left.
    \begin{array}{l}
       \xi(x,y) = x-\sum_{k=1}^N 
                          m_k\,\frac{\Delta x_k}{r_k^2},\\
       \vspace{-1ex}\\
       \eta(x,y) = y-\sum_{k=1}^N 
                          m_k\,\frac{\Delta y_k}{r_k^2},\\
    \end{array}
    \right\} \label{xi,eta:def}
 \end{equation}
 where:
 \begin{equation}
    \left.
    \begin{array}{l}
        \Delta x_k = x-x_k=r_k\cos\varphi_k,\\
        \Delta y_k = y-y_k=r_k\sin\varphi_k,\\
    \end{array}
    \right\}
 \end{equation}
 and $r_k$ and $\varphi_k$ are the corresponding polar coordinates. The
mapping functions $\xi(x,y)$ and $\eta(x,y)$ are defined everywhere in
the lens plane, except at the lensing points themselves. We shall
exclude these singular points and consider only the case $r_k\neq0$.

The microlensing mapping (\ref{xi,eta:def}) can also be regarded as a
single complex function:
 \begin{equation}
    \zeta(z) = z - \sum_{k=1}^N m_k\,\frac{z-z_k}{|z-z_k|^2} =
               z - \sum_{k=1}^N \frac{m_k}{\bar{z}-\bar{z}_k}.
    \label{zeta:def}
 \end{equation}
 Clearly, $\zeta(z)$ is not an analytic function (i.e.~it is not
complex-differentiable), since it depends on both $z$ and
$\bar{z}$. However, one can always differentiate this function {\it
partially} with respect to $x$ and $y$ (or even $z$ and $\bar{z}$).

\subsection*{Partial derivatives}

Starting from Equation~(\ref{zeta:def}) it is easy to calculate
the first partial derivatives of $\zeta(z)$ with respect to $z$ and
$\bar z$:
 \begin{equation}
   \frac{\partial\zeta}{\partial z} = 1, \hspace{8mm}
   \frac{\partial\zeta}{\partial\bar{z}} =
      \sum_{k=1}^N \frac{m_k}{(\bar{z}-\bar{z}_k)^2}.
         \label{dzeta/dz,dzeta/dzbar}
 \end{equation}
 Using the relative polar coordinates $r_k$ and $\varphi_k$, the first
derivatives can be expressed as:
 \begin{equation}
   \frac{\partial\zeta}{\partial z} = 1, \hspace{8mm}
   \frac{\partial\zeta}{\partial\bar{z}} = C_2+iS_2,
      \label{C2+iS2}
 \end{equation}
 where:
 \begin{equation}
    C_n = \sum_{k=1}^N
                  m_k\frac{\cos n\varphi_k}{r_k^n}, \hspace{8mm}
    S_n = \sum_{k=1}^N
                  m_k\frac{\sin n\varphi_k}{r_k^n}.
 \end{equation}
 From Equations~(\ref{C2+iS2}) one can also calculate the derivatives with
respect to $x$ and $y$:
 \begin{equation}
    \frac{\partial\zeta}{\partial x} = (1 + C_2) + iS_2, \hspace{8mm}
    \frac{\partial\zeta}{\partial y} = S_2 + i(1-C_2).
      \label{dzeta/dx,dzeta/dy}
 \end{equation}

Higher derivatives of $\zeta(z)$ can be obtained by further
differentiating the first
derivatives~(\ref{dzeta/dz,dzeta/dzbar}). Since $\partial\zeta/\partial
z=1$, all higher derivatives with respect to $z$ are zero. Any
mixed derivatives with respect to both $z$ and $\bar{z}$
are equal to zero as well, since $\partial\zeta/\partial\bar{z}$ does
not depend on $z$. The only remaining case involves higher derivatives
with respect to $\bar{z}$ alone:
 \begin{equation}
    \frac{\partial^s\zeta}{\partial\bar{z}^s} = (-1)^{s+1} s!
      \sum_{k=1}^N \frac{m_k}{(\bar{z}-\bar{z}_k)^{s+1}}.
 \end{equation}
 This can be rewritten in terms of the relative polar coordinates $r_k$
and $\varphi_k$ as:
 \begin{equation}
    \frac{\partial^s\zeta}{\partial\bar{z}^s} = 
          (-1)^{s+1} s! (C_{s+1}+iS_{s+1}).
            \label{d^s(zeta)/dzbar^s}
 \end{equation}
 One can also obtain the following
expression for higher derivatives with respect to $x$ and $y$:
 \begin{eqnarray}
   \frac{\partial^s\zeta}{\partial x^p\partial y^q} & = &
         (-1)^q i^q\frac{\partial^s\zeta}{\partial\bar{z}^s} \nonumber \\
    & = & (-1)^{p+1}i^q s!(C_{s+1}+iS_{s+1}),
             \label{d^s(zeta)/dx^p/dx^q}
 \end{eqnarray}
 where $s=p+q$ and $s>1$.

\subsection*{Stationary points of the mapping functions}

In order to determine whether the microlensing mapping functions
$\xi(x,y)$ and $\eta(x,y)$ have any local extrema, one has to locate the
{\it stationary points}, where the first derivatives are equal to
zero. By extracting the real and imaginary parts of
Equations~(\ref{dzeta/dx,dzeta/dy}), one obtains:
 \begin{equation}
    \left.
    \begin{array}{rcl}
       \frac{\partial\xi}{\partial x} &=& 1+C_2,\\
       \vspace{-1ex}\\
       \frac{\partial\xi}{\partial y} &=& S_2 \\
    \end{array}
    \hspace{2em}
    \begin{array}{rcl}
       \frac{\partial\eta}{\partial x} &=& S_2,\\
       \vspace{-1ex}\\
       \frac{\partial\eta}{\partial y} &=& 1-C_2.\\
    \end{array}
    \right\}
 \end{equation}
 The stationary points are found by simultaneously solving
$S_2=0$ and $C_2=-1$ for $\xi(x,y)$, and $S_2=0$ and $C_2=1$ for
$\eta(x,y)$. Obviously, the two functions cannot have any common
stationary points.

It is well known that the nature of a stationary point of a real
function $f(x,y)$ can be determined by examining the sign of the
Hessian determinant:
 \begin{equation}
    D = 
    \frac{\partial^2 f}{\partial x^2}\,
           \frac{\partial^2 f}{\partial y^2} -
           \left(\frac{\partial^2 f}{\partial x\partial y}\right)^2.
 \end{equation}
 Using Equation~(\ref{d^s(zeta)/dx^p/dx^q}) for $s=2$ and then
separating the real and imaginary parts, one can
demonstrate that both functions $\xi(x,y)$ and $\eta(x,y)$ give the same 
value for $D$:
 \begin{equation}
    D = -4(C_3^2+S_3^2).
 \end{equation}
 This excludes any possibility of $D$ being
positive, but one might still have $D=0$, if
all second derivatives are equal to zero. This means that further
examination is needed to determine the type of a stationary point.

\subsection*{Taylor's formula}
\label{taylor}

One can use Taylor's formula to express the values of the mapping
functions at a given point $\bmath{L}(x,y)$ with respect to
a nearby fixed point $\bmath{L_0}(x_{0},y_{0})$:
 \begin{equation}
  \left.
  \begin{array}{l}
   \xi(\bmath{L}) = \xi(\bmath{L_0}) + \sum_{s=1}^{n-1}
                 \frac{d^s\xi(\bmath{L_0})}{s!} +
                 \frac{d^n\xi(G)}{n!},\\
   \vspace{-1ex}\\
   \eta(\bmath{L}) = \eta(\bmath{L_0}) + \sum_{s=1}^{n-1}
                 \frac{d^s\eta(\bmath{L_0})}{s!} +
                 \frac{d^n\eta(H)}{n!},
  \end{array}
 \right\}
 \end{equation}
 where the differential operator $d^s$ is defined as:
 \begin{equation}
   d^s = \left(
                 \Delta x\frac{\partial}{\partial x} +
                 \Delta y\frac{\partial}{\partial y}
         \right)^s.
 \end{equation}
 The reminders $d^n\xi$ and $d^n\eta$ are evaluated at two different
points, $\bmath{G}$ and $\bmath{H}$, located somewhere along a section
of the straight line connecting $\bmath{L_0}$ and $\bmath{L}$.

Let us assume that $\bmath{L_0}$ is a stationary point for one of the
mapping functions. Since the first partial derivatives are zero at
$\bmath{L_0}$, the summation in Taylor's formula can effectively start
from $s=2$. In addition, we are only interested in the `problematic'
case when all second derivatives are also equal to zero. This means that
the summation can actually start from $s=3$.  In general, we have to
allow a possibility that all derivatives up to the order $(n-1)$ at
$\bmath{L_0}$ are equal to zero. However, there {\it must} exist a
high enough value for $n$ so that at least one of the $n$-th partial
derivatives is not zero. This means that either $C_{n+1}(\bmath{L_0})$
or $S_{n+1}(\bmath{L_0})$ is not zero. Taylor's formula can be
rewritten in the following form, by keeping the reminders only:
 \begin{equation}
  \left.
  \begin{array}{l}
   \Delta\xi = \xi(\bmath{L}) -\xi(\bmath{L_0}) =
          \frac{d^n\xi(\bmath{G})}{n!} =
          \frac{1}{n!}\,\mbox{\frakfamily Re}\,\delta^n\zeta(\bmath{G}),\\ 
   \vspace{-1ex}\\
   \Delta\eta = \eta(\bmath{L}) -\eta(\bmath{L_0}) =
          \frac{d^n\eta(\bmath{H})}{n!} =
          \frac{1}{n!}\,\mbox{\frakfamily Im}\,\delta^n\zeta(\bmath{H}),\\ 
  \end{array}
 \right\}
 \end{equation}
 where $\delta^n\zeta$ is the complex differential operator:
 \begin{equation}
   \delta^n = \left(
                 \Delta z\frac{\partial}{\partial z} +
                 \Delta\bar{z}\frac{\partial}{\partial\bar{z}}
         \right)^n
 \end{equation}
 It can be shown that:
 \begin{equation}
   \delta^n\zeta = \gamma\,n!\,(C_{n+1}+iS_{n+1})\,
                               (\cos n\psi-i\sin n\psi),
 \end{equation}
 where $\gamma=(-1)^{n+1}\rho^n$, $\rho$ and $\psi$ are the polar
coordinates of $\Delta\bar{z}$ and $C_{n+1}$ and $S_{n+1}$ are evaluated
either at $\bmath{G}$ (for $\Delta\xi$), or at $\bmath{H}$ (for
$\Delta\eta$). By extracting the real and imaginary parts of
$\delta^n\zeta$, one can obtain the following expressions for the
residuals $\Delta\xi$ and $\Delta\eta$:
 \begin{equation}
   \left.
   \begin{array}{l}
      \Delta\xi = \gamma\,
                    \big[C_{n+1}(\bmath{G})\cos n\psi+
                         S_{n+1}(\bmath{G})\sin n\psi\big],\\
      \vspace{-1ex}\\
      \Delta\eta = \gamma\,
                    \big[S_{n+1}(\bmath{H})\cos n\psi-
                         C_{n+1}(\bmath{H})\sin n\psi\big].
   \end{array}
   \right\}
 \end{equation}

If a mapping function has a local extremum at $\bmath{L_0}$, then the
corresponding residual must have the same constant sign at every point
$\bmath{L}$ around $\bmath{L_0}$.  It is easy to demonstrate that
neither $\Delta\xi$ nor $\Delta\eta$ can keep the same sign everywhere
around $\bmath{L_0}$. For example, let us consider $\Delta\xi$. We know
that at least one of the coefficients $C_{n+1}$ and $S_{n+1}$ is not
zero at $\bmath{L_0}$. Let $C_{n+1}(\bmath{L_0})\ne0$.  One can always
pick a point $\bmath{L_{1}}$ at $\psi_1=0$, so that $\Delta\xi_1 =
\gamma\,C_{n+1}(\bmath{G_1})$. Another point, $\bmath{L_{2}}$, can be
at $\psi_2=180\degr/n$, so that $\Delta\xi_2 =
-\gamma\,C_{n+1}(\bmath{G_2})$.  We can always choose the radius
$\rho$ to be small enough, so that $C_{n+1}(\bmath{G_1})$ and
$C_{n+1}(\bmath{G_2})$ are of the {\it same sign} depending on the
sign of $C_{n+1}(\bmath{L_0})$, since $C_{n+1}$ is continuous at
$\bmath{L_0}$. In such a case, we have opposite signs for $\Delta\xi_1$
and $\Delta\xi_2$, so that $\bmath{L_0}$ cannot be a local extremum.
If, on the other hand, we start with $S_{n+1}(\bmath{L_0})\ne0$, then
$\bmath{L_{1}}$ and $\bmath{L_{2}}$ can be set at $\psi_1=90\degr/n$ and
$\psi_2=270\degr/n$, respectively, so that the cosine function is zero
and $\Delta\xi_1$ and $\Delta\xi_2$ have opposite signs again, for the
same sign of $S_{n+1}(\bmath{G_1})$ and $S_{n+1}(\bmath{G_2})$.

Using the same idea, it is easy to demonstrate that $\Delta\eta$ cannot
be of the same sign everywhere around $\bmath{L_0}$ either. Therefore, the
microlensing mapping functions cannot have any local extrema.

\end{document}